Hamdy Elshehaby,[1,†] Omar Bakheet,[1,†] Mohamed A. Swillam[1], and Mohamed Elkabbash[2,3,*]

# Volumetric Evanescent Edge Coupling for Fiber-to-Chip Optical I/O

[1]*Department of Physics, School of Sciences and Engineering, The American University in Cairo, New Cairo 11835, Egypt*
[2]*James C. Wyant College of Optical Sciences, University of Arizona, Tucson, Arizona, USA*
[3]*Department of Physics, University of Arizona, Tucson, Arizona, USA*

[†]*These authors contributed equally.*
Email: *melkabbash@arizona.edu

***Abstract***
**Scaling optical input/output for co-packaged optics is limited by the fiber-to-chip interface. Conventional edge coupling offers low loss and broad bandwidth but confines channels to a single row along the chip facet. On the other hand, surface couplers are either narrowband or difficult to fabricate with high yield and, importantly, compete with back-end metal routing. In this work, we introduce volumetric edge coupling, in which the chip edge is structured in three dimensions so that optical coupling can occur over a two-dimensional region rather than along a single line, providing a route toward interfacing multiple waveguides with multiple cores of a multicore fiber while preserving optical access from the chip perimeter. We investigate a single-channel realization based on total-internal-reflection (TIR)-mediated evanescent coupling through the 54.7° sidewall of a KOH-etched silicon cavity, with the coupling profile shaped by a wedged buried-oxide ridge. Finite-difference time-domain simulations predict a peak coupling efficiency of 88% (−0.56 dB) at 1550 nm and a 1-dB bandwidth of 86.45 nm spanning the C-band. The design further exhibits a vertical alignment tolerance of approximately ±2 µm, weak sensitivity to transverse offsets up to 5 µm, and a ±0.8° 1-dB angular tolerance. The reflected optical field also provides a potential alignment signal, offering a path toward reduced active alignment overhead in future multicore implementations.**

## Introduction

Modern AI data centers increasingly rely on optical interconnects to support the rapidly growing bandwidth requirements of data movement between switches and computing resources. In conventional architectures, pluggable optical transceivers are located at the front panel of the switch, requiring high-speed electrical signals to travel several centimeters between the switch application-specific integrated circuit (ASIC) and the optical modules. At aggregate switch bandwidths approaching tens of terabits per second, these electrical links contribute appreciable energy consumption and are increasingly susceptible to signal-integrity limitations, including electromagnetic crosstalk [1]. Co-packaged optics (CPO) addresses these limitations by placing the optical engines within the same package as the switch ASIC, thereby reducing the length of the high-speed electrical interconnects from centimeters to millimeters[2].

Many proposed CPO architectures employ external laser sources to reduce the thermal and integration challenges associated with placing semiconductor lasers in close proximity to high-power switching electronics. Semiconductor laser emission is temperature dependent, with wavelength shifts on the order of 0.1 nm/°C reported for typical devices[3]. This sensitivity can complicate wavelength stabilization in wavelength-division-multiplexed systems when the laser source is exposed to the thermal environment of the switch package. External-laser architectures therefore retain an optical connection between the light source and the photonic integrated circuit, making efficient and scalable fiber-to-chip coupling an important component of CPO packaging.

Conventional fiber-to-chip interfaces commonly rely on edge coupling, in which light is injected through the chip facet and an inverse taper or other mode-conversion structure is used to improve the overlap between the fiber mode and the on-chip waveguide mode[4]. Edge couplers can provide low coupling loss and broad optical bandwidth; however, their scaling is geometrically constrained because the fibers are typically arranged along a single row at the chip perimeter. Standard V-groove fiber arrays commonly employ center-to-center fiber pitches of 127 or 250 µm, corresponding to only several optical channels per millimeter of chip edge. As optical I/O requirements increase toward hundreds or potentially thousands of channels per package, this one-dimensional arrangement becomes an increasingly important limitation on edge-coupled optical connectivity [5].

Surface coupling provides an alternative route in which optical access is provided through the top surface of the chip rather than through its perimeter. This allows coupling sites to be distributed over the two-dimensional chip area rather than along a single edge. Surface-coupling approaches include grating couplers [6], elephant couplers [7], and integrated micro-mirror couplers[8]. Among these approaches, grating couplers are particularly attractive because they can be fabricated using planar lithographic processes and have therefore been widely adopted in silicon-photonics research and packaging. Their periodic nature, however, introduces wavelength-dependent coupling conditions that can limit the usable optical bandwidth unless broadband designs are specifically engineered [9]. In addition, surface coupling requires optical access to the chip surface and therefore competes with back-end-of-line metallization, electrical routing, thermal-control structures, and other functional elements for available chip area. Dedicated optical-access windows through the back-end stack may therefore be required. Edge coupling, in contrast, confines the optical access region primarily to the chip perimeter and can leave the top surface available for electrical and thermal integration.

The resulting question is whether a perimeter-based optical interface can overcome the geometric scaling limitation of conventional edge coupling while preserving the broadband and low-loss characteristics that make edge access attractive. Here, we introduce volumetric edge coupling, an architecture that extends the optical coupling region from a one-dimensional arrangement along the chip facet into a two-dimensional region formed within a three-dimensionally structured chip edge, as illustrated in Figure 1. In conventional edge coupling, shown in Figure 1(a), the optical channels are arranged along a single line at the chip facet, so the achievable channel density is constrained by the fiber pitch along the perimeter. In volumetric edge coupling, shown in Figure 1(b), the chip edge is instead structured in three dimensions, creating a coupling region within which coupling sites could be distributed both laterally and vertically. This architecture provides a route toward interfacing multiple waveguides with multiple cores of a multicore fiber while retaining optical access from the chip perimeter. In this work, we establish the operating principle through the design and numerical demonstration of a single coupling channel, which serves as the building block for future multichannel implementations.

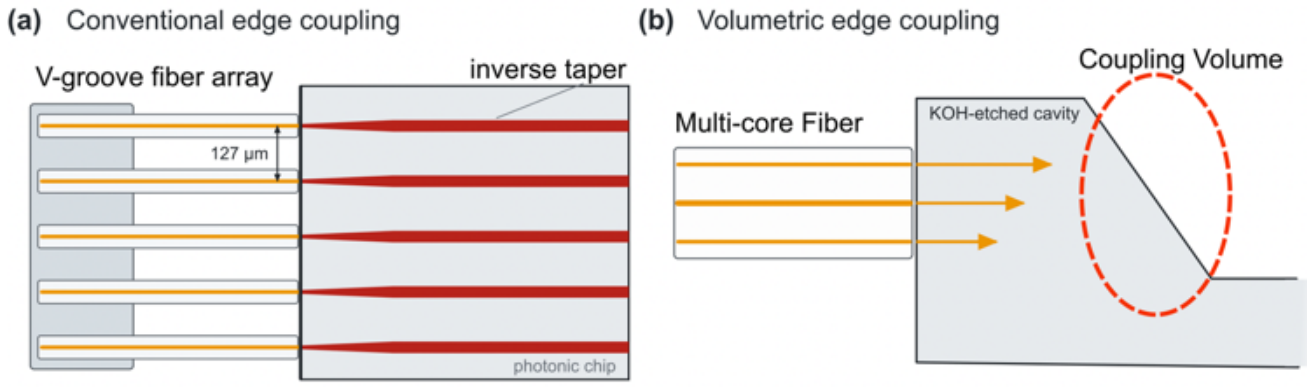

Figure 1 Concept of volumetric edge coupling. (a) In a conventional edge coupler, optical channels are arranged along a single line at the chip facet, constraining the channel count by the one-dimensional fiber pitch along the perimeter. (b) Volumetric edge coupling structures the chip edge in three dimensions, creating a two-dimensional coupling region within which multiple coupling sites can be distributed. The illustrated multicore-fiber interface represents the envisioned multichannel extension of the single-channel coupling mechanism demonstrated in this work.

## OPERATION PRINCIPLE:

The proposed coupler operates through evanescent coupling mediated by total internal reflection (TIR) at an anisotropically etched silicon sidewall. Anisotropic KOH etching of a (100) silicon substrate exposes slow-etching {111} facets inclined by approximately 54.74°with respect to the wafer surface. A TE-polarized Gaussian beam launched toward the etched cavity is incident on the silicon–oxide interface at an angle exceeding the critical angle for TIR and is consequently reflected into the silicon substrate, as illustrated in Figure 2(a). Although the incident field undergoes TIR, an evanescent field extends beyond the interface into the dielectric region. A silicon waveguide positioned within this evanescent field interacts with the incident optical field, enabling power to be transferred into its guided mode. The coupling efficiency is governed by the propagation characteristics of the interacting fields and their spatial overlap along the coupling region.

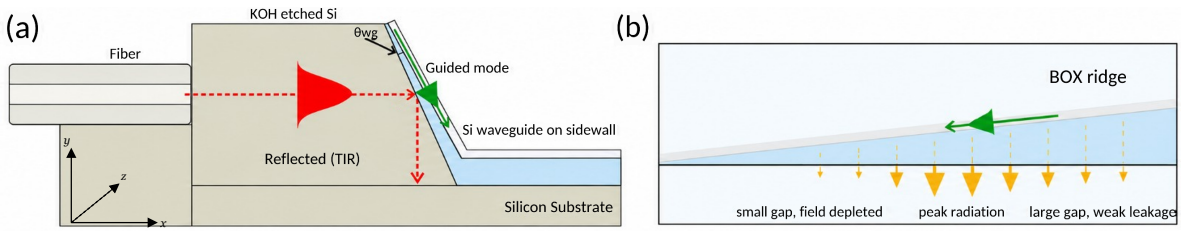

Figure 2 Proposed fiber-to-chip coupler based on volumetric edge coupling. (a) Schematic of the coupling structure, showing the anisotropically etched silicon cavity with its 54.7° sidewall, the incident Gaussian beam undergoing total internal reflection at the sidewall, and the silicon waveguide patterned on the wedged buried oxide that picks up the evanescent field. (b) Apodization mechanism produced by the angled BOX ridge: because the silicon–sidewall gap varies along the propagation direction, the local coupling strength is shaped along the interaction length so that, in the time-reversed picture, the outcoupled field builds a Gaussian envelope matched to the fiber mode.

Efficient transfer into the waveguide requires control over both the guided-mode properties and the spatial distribution of the coupling strength. The waveguide thickness ($t_{wg}$) determines the confinement and spatial extent of the guided mode and therefore influences its interaction with the evanescent field. In the present implementation, the waveguide is assumed to be silicon; however, the volumetric coupling concept is not fundamentally restricted to crystalline-silicon waveguides, and deposited waveguide materials could be considered in future implementations.

The second geometric degree of freedom is the dielectric separation between the waveguide and the TIR interface. Here, the dielectric spacer is assumed to be the buried oxide (BOX) layer. A BOX ridge is introduced beneath the coupling region and oriented at an angle relative to the KOH-etched sidewall, causing the perpendicular separation between the waveguide and the sidewall to vary continuously along the propagation direction, as shown in Figure 2(b). Because the evanescent field decays with distance from the TIR interface, this varying separation produces a spatially varying interaction strength between the incident field and the waveguide. The BOX geometry therefore provides a means of shaping the coupling profile along the interaction length, analogous to the spatial apodization employed in grating couplers[10].

The role of this spatially varying coupling can be understood intuitively by considering the time-reversed process, in which the guided mode is progressively coupled out of the waveguide toward the incident optical mode. Near the beginning of the interaction region, the relatively large waveguide–sidewall separation produces weak coupling, so only a small fraction of the guided power is extracted. As the separation decreases along with the propagation direction, the interaction becomes stronger and a larger fraction of the remaining guided power is transferred from the waveguide. Toward the end of the coupling region, the interaction is strongest, but less guided power remains available for extraction. The resulting longitudinal coupling profile can therefore be engineered through the BOX geometry to shape the outcoupled field toward the Gaussian spatial envelope of the fiber mode, thereby improving the mode overlap between the two fields.

The continuously wedged BOX considered here provides a convenient model for demonstrating this coupling principle, although realizing such a geometry may present fabrication challenges. More fabrication-compatible implementations could reproduce a similar spatial variation in coupling strength by engineering the effective refractive index of the dielectric spacer, for example using compositionally graded silicon oxynitride [11] or through nanostructing the waveguide layer to spatially control the mode confinement and, subsequently, the spatially varying evanescent coupling [12].

## Geometric Optimization and Performance Analysis

To optimize the proposed coupler, two-dimensional finite-difference time-domain (2D-FDTD) simulations were performed using Ansys Lumerical®. Perfectly matched layer (PML) boundary conditions were applied on all sides of the simulation domain to suppress reflections from the simulation boundaries. The input source was modeled as a Gaussian beam with a waist radius of 5 µm, representing the mode profile of a single-mode fiber. In the 2D coordinate system, the source propagates along the x-axis, with its field profile distributed along the y-axis. The vertical source position ($y$) was referenced to the bottom of the etched silicon cavity and included as an optimization parameter to determine the position providing maximum coupling to the guided waveguide mode. A frequency-domain power monitor and a mode-expansion monitor were used to evaluate the optical power transmitted into the waveguide and determine the coupling efficiency.

The geometrical optimization was initially performed in 2D by sweeping the waveguide thickness ($t_{wg}$), source position ($y$), and waveguide angle ($\theta_{wg}$). Following the 2D optimization, three-dimensional FDTD simulations were performed for different waveguide widths to evaluate the influence of the finite lateral dimension neglected in the 2D model. As the waveguide width increased, both the coupling loss and spectral bandwidth approached the corresponding 2D results. This behavior is expected because increasing the width reduces the influence of the lateral boundaries, causing the 3D structure to progressively approximate the laterally invariant geometry represented by the 2D model. Beyond this regime, further increases in the waveguide width produced only minor changes in the coupling performance. The computationally less demanding 2D model was therefore used for the subsequent optimization and tolerance analyses involving parameters contained within the 2D simulation plane, while the effect of out-of-plane displacement was evaluated using the full 3D model.

Figure 3 compares the normalized guided-mode transmission spectra obtained from the 2D simulation with those obtained from the 3D simulations for different waveguide widths. The optimized geometry has a waveguide thickness $t_{wg}$ = 262 nm, a waveguide angle $\theta_{wg}$ = 53.50°, and a source position $y$ = 11 µm relative to the cavity base. As summarized in Table 1, this configuration yields a maximum normalized guided-mode transmission of 0.88, corresponding to a coupling loss of 0.56 dB, at a center wavelength of 1550 nm. The coupling efficiency is evaluated in the angled section of the waveguide immediately following the coupling region and before the subsequent waveguide bend. Therefore, the reported coupling efficiency characterizes the proposed fiber-to-waveguide coupling mechanism and does not include additional scattering loss that may arise from the transition between the angled and straight waveguide sections

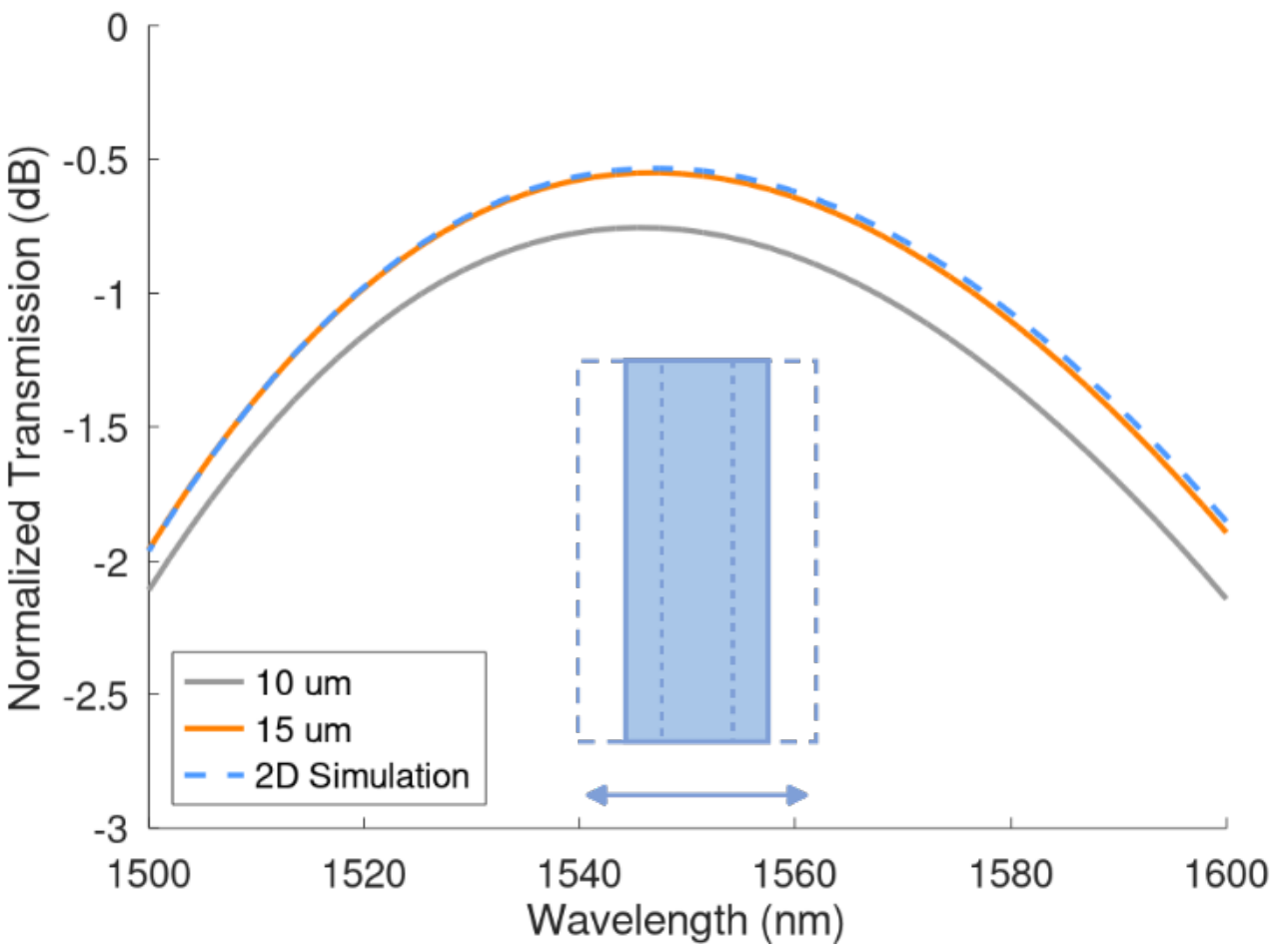

Figure 3 Normalized transmission spectra for 2D simulation and different waveguide widths in 3D simulation.

Table 1. Optimized Geometric Parameters and Peak Coupling Performance

| Parameter | Value |
|---|---|
| Source Position ($y$) | 11 $\mu m$ |
| Waveguide Angle ($\theta_{wg}$) | 53.50° |
| Waveguide Thickness ($t_{wg}$) | 262 $nm$ |
| $\lambda_{max}$ | 1550 $nm$ |
| $T_{max}$ | -0.56 dB |

## Spectral Bandwidth and Robustness

The spectral performance of the optimized structure is summarized in Table 2. The coupler exhibits a 1-dB bandwidth of 86.45 nm, extending from 1507.23 nm to 1593.68 nm and covering the entire C-band. The broad spectral response observed in Figure 3 can be understood from two main characteristics of the coupling mechanism.

First, the proposed structure does not rely on a resonant cavity or a periodic perturbation to mediate power transfer. In conventional grating couplers, the periodic structure introduces an additional momentum component determined by the grating period, resulting in a wavelength-dependent phase-matching condition. In the proposed coupler, power transfer instead arises from the interaction between the evanescent field generated by total internal reflection at the KOH-etched sidewall and the nearby guided waveguide mode. The absence of a periodic perturbation eliminates the wavelength dependence associated with a fixed grating momentum, contributing to the broad spectral response observed in the simulations.

The bandwidth is nevertheless limited by the wavelength dependence of the interaction between the evanescent field and the guided waveguide mode. The penetration depth of the evanescent field generated at the TIR interface varies with wavelength, modifying the field available for coupling at a given waveguide–sidewall separation. At the same time, the confinement and spatial extent of the guided mode also vary with wavelength. Consequently, the longitudinal coupling profile produced by the varying waveguide–sidewall separation, which is optimized at the design wavelength of 1550 nm, changes as the wavelength is detuned. In the time-reversed picture illustrated in Figure 2(b), these variations correspond to a wavelength-dependent change in the spatial distribution of the outcoupled field and consequently its overlap with the Gaussian fiber mode. These wavelength-dependent changes in coupling strength and spatial mode overlap ultimately limit the bandwidth of the device.

TABLE 2. CALCULATED SPECTRAL BANDWIDTH PERFORMANCE FOR THE OPTIMIZED COUPLER

| Parameter | -1 dB |
|---|---|
| $\lambda_{low}$ $(nm)$ | 1507.23 |
| $\lambda_{high}$ $(nm)$ | 1593.68 |
| Bandwidth $(nm)$ | 86.45 |

## FABRICATION AND ALIGNMENT SENSITIVITY ANALYSIS

To evaluate the practical viability of the proposed fiber-to-chip coupler, we performed tolerance analysis considering both fabrication variations and source misalignment during packaging. The fabrication and alignment tolerances were quantified in terms of the allowable variation in each parameter that results in a maximum 1-dB reduction in coupling efficiency relative to the optimized value at 1550 nm. The parameter values and their corresponding 1-dB tolerances are summarized in Table 3.

To evaluate the robustness of the proposed coupler against fabrication variations, a fabrication tolerance analysis was performed for the two main geometric parameters of the coupling structure: the waveguide angle ($\theta_{wg}$) and the waveguide thickness ($t_{wg}$). The sensitivity of the coupling efficiency to variations in these parameters is shown in Figure 4. The optimized waveguide angle of 53.5° exhibits a tolerance of approximately ±0.4°. Variations in the waveguide angle modify the rate at which the separation between the waveguide and the KOH-etched sidewall changes along the propagation direction. This alters the spatial distribution of the evanescent coupling strength and, consequently, the effective apodization of the coupling region. As the angle deviates from its optimum value, the resulting coupling profile becomes less well matched to the incident Gaussian field, leading to a gradual reduction in coupling efficiency. The optimized waveguide thickness of 262 nm exhibits a tolerance of approximately ±7 nm. Changes in thickness modify the confinement and spatial extent of the guided mode, thereby changing its overlap with the evanescent field generated at the TIR interface. Thickness variations can also modify the propagation characteristics of the guided mode, further affecting the interaction between the two fields. The observed tolerances therefore reflect the combined sensitivity of the coupler to the spatial coupling profile and the modal properties of the silicon waveguide.

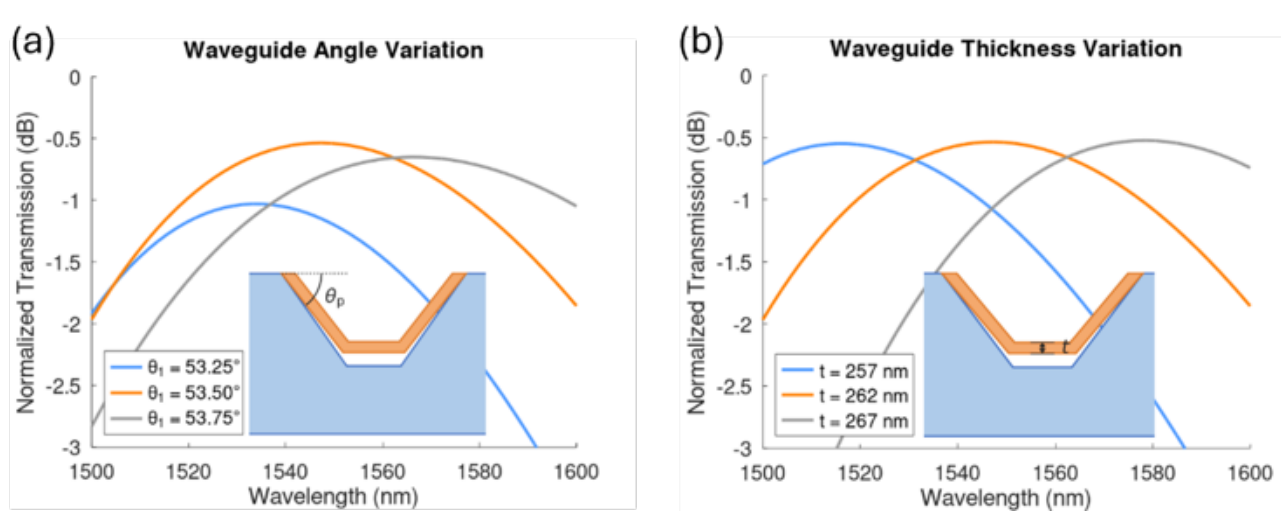

Figure 4 Fabrication tolerance analysis of the proposed on-chip coupler. (a) Effect of waveguide angle ($\theta_{wg}$) variations. (b) Effect of waveguide thickness ($t_{wg}$) variations.

In addition to fabrication variations, the sensitivity of the proposed coupler to source misalignment was investigated to evaluate its suitability for practical fiber-to-chip packaging. The coupling efficiency was analyzed as a function of vertical source displacement, source tilt angle, and transverse source displacement, as shown in Figure 5. The vertical-displacement and tilt-angle analyses were performed using the 2D model, since both variations occur within the simulation plane. In contrast, transverse displacement occurs along the out-of-plane direction and therefore requires the full 3D model.

As shown in Figure 5(a), the coupler exhibits a 1-dB vertical displacement tolerance of approximately ±2 μm at 1550 nm. Vertical displacement changes the position of the incident Gaussian beam relative to the coupling region and therefore modifies its spatial overlap with the coupling profile. As the source moves away from its optimum position, this overlap deteriorates, resulting in a gradual reduction in coupling efficiency. The obtained ±2 μm tolerance is larger than that of many conventional inverse-taper edge-coupling interfaces, which commonly require sub-micrometer to approximately micrometer-scale fiber positioning, and is comparable to the more alignment-tolerant edge-coupler architectures reported in the literature [13].

The effect of transverse displacement is presented in Figure 5(b). Since transverse displacement occurs along the out-of-plane direction, this analysis was performed using the full 3D model. Transverse offsets of up to 5 μm produce only a small variation in transmission, demonstrating substantially weaker sensitivity to lateral misalignment than is typical of many conventional fiber-to-chip edge-coupling interfaces. This robustness arises from the comparatively large lateral extent of the coupling region, which allows strong spatial overlap between the incident field and the supported waveguide mode to be maintained as the source is displaced transversely. The simulated response is symmetric with respect to positive and negative transverse displacement; therefore, only the positive displacement range is shown. The transverse alignment tolerance is superior to that of edge couplers [13].

The sensitivity to angular misalignment is shown in Figure 5(c). The source exhibits a 1-dB tilt tolerance of approximately ±0.8°, making tilt the most sensitive of the investigated alignment parameters. Tilting the incident beam changes both its propagation direction at the TIR interface and its spatial trajectory across the coupling region. Consequently, the distribution of the evanescent field along the sidewall and its interaction with the guided mode are modified, leading to a reduction in coupling efficiency as the tilt deviates from its nominal value. The obtained angular tolerance is comparable to values reported for practical fiber-to-chip interfaces. For example, an arrayed Silicon Nitride grating coupler interface demonstrated a measured 1-dB angular tolerance of approximately ±1°[14]. Other micro-optical coupling architectures have demonstrated angular tolerances on the order of a few degrees [15], emphasizing that angular sensitivity depends strongly on the coupling geometry, beam size, and rotational degree of freedom being considered.

Although angular alignment represents the most restrictive degree of freedom in the present design, its impact on the envisioned multichannel implementation is mitigated by the architecture of a multicore fiber. Since the individual cores share a common fiber axis, fiber tilt constitutes a global alignment degree of freedom rather than an independent parameter for each optical channel. The angular orientation can therefore be optimized for the complete multicore-fiber assembly rather than independently for every core.

The relatively tight angular tolerance can be further addressed at the packaging level by exploiting the reflected optical field as an alignment signal. The evanescent coupling mechanism inherently links the power transferred to the guided mode to the optical power remaining in the reflected field. In the absence of efficient coupling, a large fraction of the incident power remains in the reflected beam following total internal reflection at the coupling interface. As the source approaches its optimum alignment, increased power transfer to the guided mode is accompanied by a reduction in reflected power. This behavior is demonstrated in Figure 5(d), where the normalized reflected power reaches its minimum near the nominal 0°source tilt, corresponding closely to the orientation at which the maximum waveguide transmission is obtained in Figure 5(c). Conversely, increasing angular misalignment results in reduced waveguide coupling and a corresponding increase in reflected power. The reflected field therefore provides a directly accessible optical observable that could be used as feedback during angular alignment.

This capability is particularly attractive for active photonic packaging. Conventional active-alignment schemes can employ additional on-chip monitoring structures, such as directional couplers and photodetectors, to determine the optical power coupled into the input waveguides during assembly[5]. Such structures introduce additional device, routing, and chip-area overhead, which can become increasingly significant as the number of optical channels increases. In the proposed architecture, the quality of the optical alignment could instead be inferred directly from the externally accessible reflected field, potentially eliminating the need for a dedicated on-chip alignment monitor for each input channel. Thus, although the intrinsic 1-dB angular tolerance remains approximately ±0.8°, reflection-assisted feedback provides a potential means of mitigating the practical packaging challenge associated with this relatively tight tolerance by enabling the optimum angular orientation to be identified directly during assembly.

This approach could become particularly advantageous in the envisioned multicore-fiber implementation. Since tilt is a global alignment parameter for the multicore fiber, reflected signals from multiple cores could potentially be monitored simultaneously using an external camera or detector array. The orientation of the complete fiber assembly could then be optimized using the reflected-power distribution without requiring an independent on-chip monitoring circuit for every optical channel.

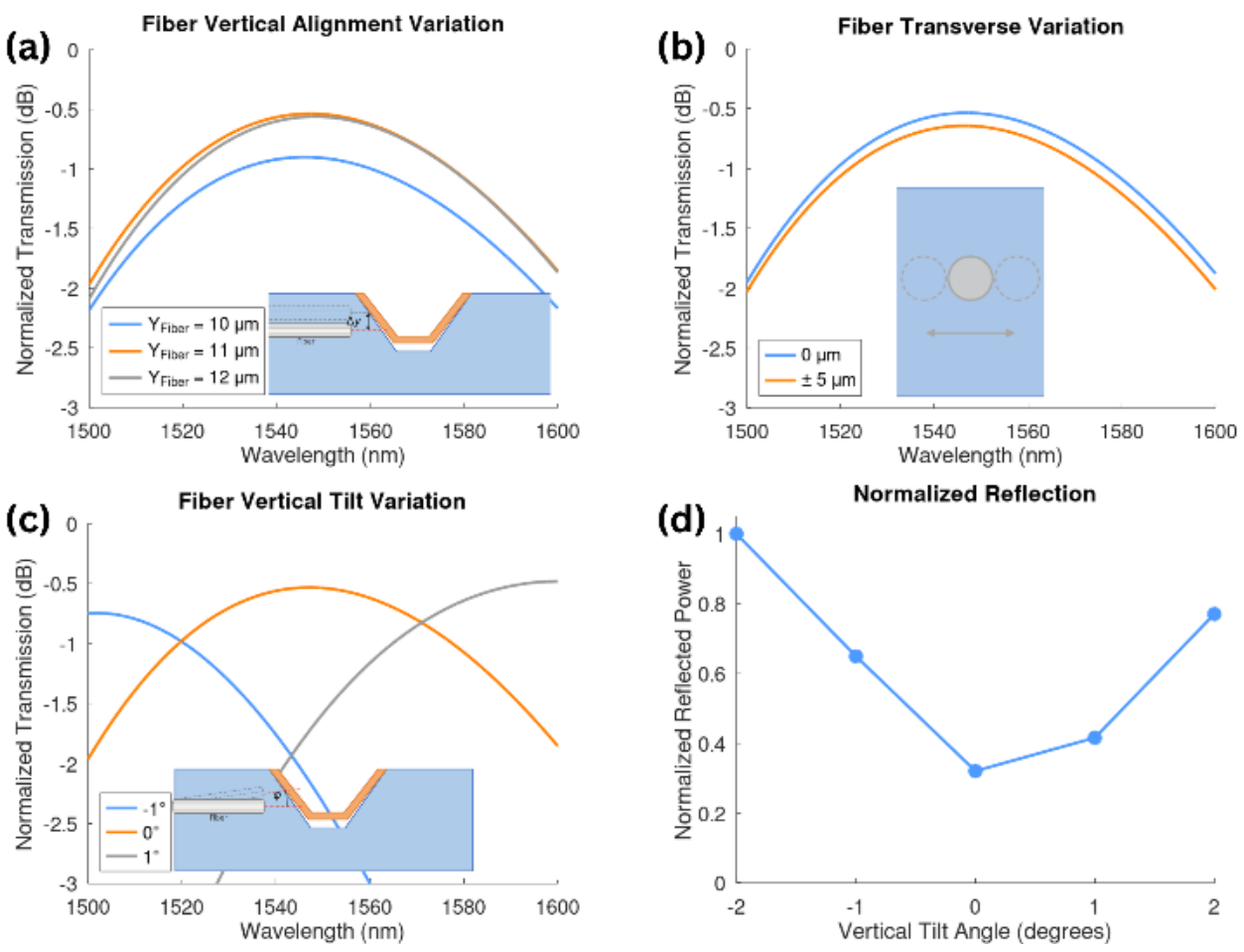

Figure 5 Alignment sensitivity analysis of the proposed on-chip coupler. (a) Sensitivity to vertical source displacement and (b) Fiber Transverse Variation. (c) Effect of source tilt angle ($\phi$) relative to the x-axis and the corresponding normalized reflection that shows an off-chip alignment method compatible with the proposed chip coupling method (d).

TABLE 3. SUMMARY OF FABRICATION AND ALIGNMENT TOLERANCES

| Parameter | Nominal | 1 dB Tolerance |
|---|---|---|
| Waveguide Angle ($\theta_{wg}$) | 53.50° | ±0.4° |
| Thickness ($t_{wg}$) | 262 $nm$ | ±7 nm |
| Source Position ($y$) | 11 $\mu m$ | ±2 $\mu m$ |
| Source Tilt Angle ($\theta_{source}$) | 0° | ±0.8° |

## CONCLUSION AND OUTLOOK:

We introduced volumetric edge coupling as an architecture for extending edge based optical I/O beyond the one-dimensional arrangement of conventional edge couplers. By structuring the chip edge in three dimensions, the proposed approach creates a coupling region in which multiple coupling sites could ultimately be distributed across the edge while leaving the chip surface available for electrical interconnects while retaining its low-loss and broadband character. In the proposed silicon implementation, a KOH-etched 54.7°sidewall generates the evanescent field, while a shaped dielectric spacer controls the local coupling strength and mode matching. The optimized design reaches 88% coupling efficiency (-0.56 dB) at 1550 nm with a 1-dB bandwidth of 86.45 nm. The design also shows practical alignment robustness, including approximately ±2 µm vertical displacement tolerance and only modest degradation for transverse offsets up to 5 µm. The most sensitive alignment parameter is fiber tilt, with a ±0.8° 1-dB tolerance, but this angle can be controlled globally for a multicore fiber or rigid fiber array rather than independently for every optical channel. Together with reflection-based monitoring enabled by the FTIR coupling mechanism, these characteristics suggest a path toward reducing the active-alignment complexity and assembly time associated with conventional fiber-to-chip packaging.

The present work is intended as a first proof of principle and evaluation step for volumetric edge coupling architecture. Future works will extend the single-channel concept to simultaneous coupling from multiple cores of a multicore fiber into multiple waveguides. This extension will require variation in the effective waveguide angle, or the corresponding index gradient, or through varying the effective modal index/modal area of the waveguide layer through introducing a spatially varying subwavelength features in the waveguide. Silicon Nitride and other deposited waveguide materials will be investigated in place of crystalline silicon so that the waveguide layer can be conformally deposited over the etched sidewall, avoiding the need for crystalline-silicon transfer or bonding while preserving the volumetric evanescent-coupling mechanism. Because the coupling efficiency reported here is evaluated in the inclined waveguide immediately after the evanescent-coupling region. The coupling efficiency reported here therefore represents the power transferred to the inclined receiving waveguide and does not account for the subsequent transition to the planar photonic layer, which will be addressed in future work.